\documentclass[showpacs,preprintnumbers,amsmath,amssymb]{revtex4}

\usepackage{graphicx}
\usepackage{dcolumn}
\usepackage{bm}

\newcommand{\D}{\displaystyle}
\begin{document}

\title{Synchronization and clustering of synthetic genetic networks: A role for cis-regulatory modules}

\author{Jiajun Zhang$^1$}
\author{Zhanjiang Yuan$^1$}
\author{Tianshou Zhou$^{2,1,}$}
\email{mcszhtsh@mail.sysu.edu.cn}

\affiliation{$^1$School of Mathematics and Computational Science,
Sun Yat-Sen University, Guangzhou 510275, China\\
$^2$State Key Laboratory of Biocontrol and Guangzhou Center for
Bioinformatics, School of Life Science, Sun Yat-Sen University,
Guangzhou 510275, China}

\date{\today}

\begin{abstract}
The effect of signal integration through cis-regulatory modules
(CRMs) on synchronization and clustering of populations of
two-component genetic oscillators coupled by quorum sensing is in
detail investigated. We find that the CRMs play an important role in
achieving synchronization and clustering. For this, we investigate 6
possible cis-regulatory input functions (CRIFs) with AND, OR, ANDN,
ORN, XOR, and EQU types of responses in two possible kinds of
cell-to-cell communications: activator-regulated communication
(i.e., the autoinducer regulates the activator) and
repressor-regulated communication (i.e., the autoinducer regulates
the repressor). Both theoretical analysis and numerical simulation
show that different CRMs drive fundamentally different cellular
patterns, such as complete synchronization, various cluster-balanced
states and several cluster-nonbalanced states.
\end{abstract}


\maketitle

\section{Introduction}

Decoupling simple networks from their native yet often complex
biological settings can lead to valuable information regarding
evolutionary design principles. This motivates the design and
construction of synthetic genetic networks resembling submodules of
natural circuitry in vivo, which in turn lead to the construction of
devices and softwares capable of performing elaborate functions in
living cells \cite{Bray_Nature95}. Due to recent advances in
bioengineering technology, several prototype synthetic genetic
motifs, such as logic gates \cite{Weiss_NaturalComput03,Anderson07},
toggle switches \cite{Gardner_Nature00,Kim06}, and oscillators
\cite{MBElowitz-Nature00,MRAtkinson-Cell03,Hasty-Chaos01} have been
successfully constructed. These simple architectures are thought of
as essential modules in living organisms, and based on them,
complementary approaches have been developed to explore the
relationship between the structure and function of more complex
genetic circuits \cite{Hasty-Nature02,Sprinzak-Nature05}.

A natural step in the design of artificial gene networks would be to
include a mechanism of intercell coupling that would globally
enhance, given that cells are frequently subject to chemical signals
from neighboring cells, the oscillating response of the system. The
most common communication mechanism with such a function is quorum
sensing, the ability of bacteria to communicate with each other
through signaling molecules that are released into the cellular
environment. Quorum sensing has lead to programmed population
control in a bacterial population \cite{You,Balagadde05,
Brenner,Basu04,Basu05,Halsintine,Balagadde08,DMcMillen-PNAS02,JGarciaOjalvo-PNAS04,AKuznetsov-SIAM04,
TSZhou-PRL05,TSZhou-PlosOne07,Ullner07,Koseska_PRE07_1,
Koseska_PRE07_2,Koseska_PRE07_3,Yuan_PhysRevE08,chaos-08,Ullner_PRE08}.
Through such a mechanism, the ability of cells to communicate to one
another allows them to coordinate the behavior of the entire
community, where gene expression is regulated in response to the
local cell population density \cite{Fuqua96}. A well-defined example
of coordinated global behavior in bacteria is a population of
genetic relaxation oscillators coupled to a quorum sensing
apparatus, which can achieve synchronization through the so-called
``fast threshold modulation" mechanism \cite{DMcMillen-PNAS02}.
Coupling, however, can be devised in different ways, e.g.,
attractive or repulsive cell-to-cell communication
\cite{chaos-08,DMcMillen-PNAS02,Ullner07,JGarciaOjalvo-PNAS04}, in
synthetic systems. Different couplings would lead to different
dynamic patterns, such as synchronization, clustering and
multistability \cite{chaos-08,Ullner07}.

Why is there such a difference in cellular patterns when different
types of cellular communication are employed? Actually, biological
functions appearing as collective behaviors may arise from a
particular module that integrates intracellular and extracellular
signals. Such a module is now known as cis-regulatory module (CRM),
which contains a cluster of binding sites for transcription factors
(TFs) and determine the place and timing of gene action within the
network, e.g., the CRM in the sea urchin embryo can control not only
static spatial assignment in development but also dynamic regulatory
patterning \cite{Smith07}. TFs are often integrated in a
combinatorial logic manner, and moreover such a combination may take
different schemes
\cite{Alon07,Buchler03,PlosComputBiol06,Mangan03,Setty03,PlosBiol06},
leading to different CRMs. In fact, from views of evolutionism, CRMs
are changeable, e.g., cis-regulatory mutations \cite{AWGregory07},
and such a mutation constitutes an important part of the genetic
basis for adaptation. However, how different CRMs affect collective
behaviors across ensembles of genetic oscillators with cell-to-cell
communication remains to be fully explored.

In this paper, we investigate this question in detail and find that
CRMs play a significant role in the mode of dynamic patterns at the
cellular population level, e.g., the CRMs can drive fundamentally
different cellular patterns such as synchronization and clustering.
We first design and construct a multicellular network with a CRM,
using a variant of the synthetic genetic relaxation oscillator
developed in {\it E.coli} \cite{Hasty-Chaos01} and utilizing quorum
sensing to communicate between cells. Since different CRMs due to
cis-regulatory mutations \cite{AWGregory07} lead to different types
of cis-regulatory input functions (CRIFs) such as AND, OR, ANDN,
ORN, XOR, EQU, we then investigate the effects of these different
CRIFs on cellular patterns to support our conclusion. We emphasize
that since the proposed genetic relaxation oscillator is composed of
interacting positive and negative feedback loops, and this circuit
topology is common in genetic oscillators such as cell cycle and
circadian clocks
\cite{Tsai08,Liu97,Mihalcescu04,Pomerening05,Rust07,Yamaguchi03,Hasty_Nature08},
our conclusion on how CRMs influence the dynamics of genetic
circuits with this shared topology will be of general relevance to a
wide range of cellular processes.

\section{Mathematical model and theoretical analysis}

\subsection{Model}

\begin{figure}
  \centering
  \includegraphics[width=10cm]{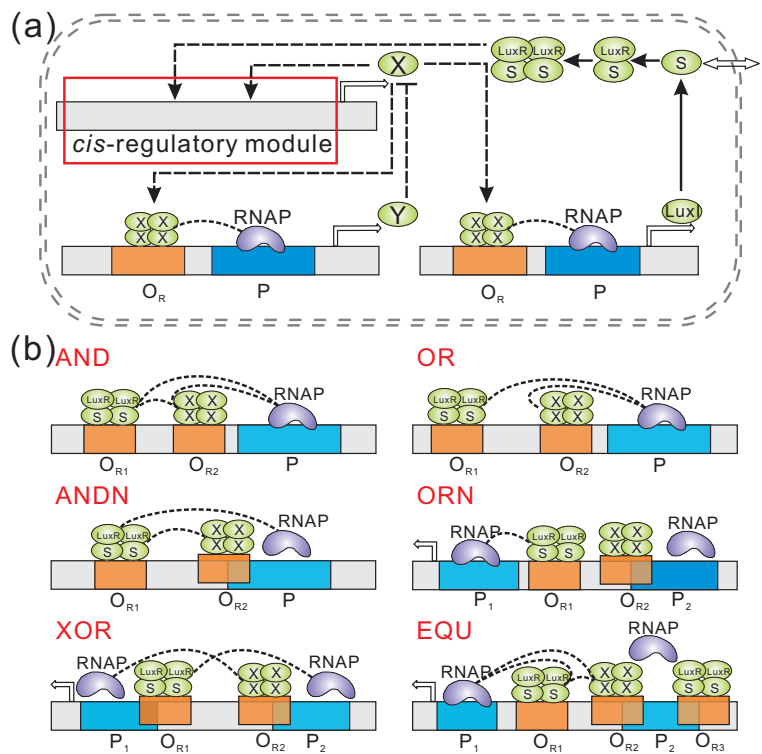}
  \caption{(Color online) (a) Schematic diagram for the network of
  genetic relaxation oscillators with a cis-regulator module, where the right bidirectional arrow indicates that S can freely
  diffusive through the cellular membrane.
  (b) Six cis-regulatory constructs for implementations of six
  different logic functions. In (a) and (b), X, Y and LuxI denote the proteins, P$_1$ and P$_2$ represent the promoters.
  O$_{\rm R}$ stands for operator site whereas RNAP for RNA polymerase.
  We use offset and
  overlapping boxes to indicate the mutual repression and the
  dashed lines to indicate the cooperative interaction.}
  \label{fig1}
\end{figure}
First, we report our design on a network of coupled synthetic
genetic relaxation oscillators with a CRM, which is schematically
shown in Fig. \ref{fig1}. The core oscillator is a variant of the
genetic relaxation oscillator proposed in Ref. \cite{Hasty-Chaos01}.
In such an oscillator, the activator X (CII) and the repressor Y are
under the control of different promoters from the $\lambda$ phage
virus. In Fig. \ref{fig1}(a), X is the autocatalytic portion of the
oscillator whereas Y is a protease that degrades X. Both genes x and
y are activated by protein X. Such a circuity not only is a useful
architecture to understand information processing of simple
oscillators but also appears as a common core motif in biological
contexts
\cite{Tsai08,Liu97,Mihalcescu04,Pomerening05,Rust07,Yamaguchi03}. In
our design, we utilize the quorum-sensing apparatus of the bacterium
{\it Vibrio Fischeri} \cite{Fuqua96} to communicate between cells.
This cell-to-cell communication system operates by diffusing a small
molecule [also called autoinducer (AI)] into the environment. Since
the communication is implemented by the signal molecule which
regulates the activator X, we refer to it as activator-regulated
communication. When this molecule binds to a regulatory protein
(LuxR), both it and X bind the regulatory region of gene x or y and
combinatorially modulate the transcription rate. Many of these
combinational effects are performed by a CRM, which can function as
analogous implementations of logic gates. The corresponding CRM
contains a cluster of binding sites of two different transcription
factors (TFs) that control the activation or repression of a gene.
These TFs may be either activators enhancing the binding or the
activity of the RNA polymerase in the cognate promoters, or
repressors blocking this binding, or both via the mechanism of
``regulated recruitment" \cite{Ptashne02}. Based on the possible
combination of the two TFs, the CRM can perform different logic
functions with different implementations, as shown in Fig.
\ref{fig1}(b). Limited by the regulatory structure of the relaxation
oscillator (more precisely, the TF X serves as activator only), we
have six biologically feasible CRM designs: AND, OR, ANDN, ORN, XOR
and EQU (see Table \textrm{I}). These logic functions have been
either described experimentally or suggested to occur on the basis
of simulations using empirical data \cite{FEBSLett08}. Actually, the
prokaryotic transcription networks provide a large number of
composite logic operators that are implemented through more complex
natural or simulated regulatory setups. Alternatively, the CRM
designs can be implemented by introducing mutations at the
amino-acid sequences of the TFs and the bp sequences of the
cis-regulatory regions \cite{PlosComputBiol06}. Note that, in our
designs, the signaling molecules can serve as not only activators
but also repressors by the introduction of an alternative promoter
\cite{Egland01}.

\begin{table}[htb] \label{t1}
\caption{Logic operations for cis-regulatory input functions}
\vspace{-0.5cm}
\[
\begin{array}{ccccccccc}
\hline\hline
\multicolumn{2}{c}{\mbox{TFs}}& & \multicolumn{6}{c}{\mbox{Logic functions}} \\
\cline{1-2}\cline{4-9}
\multicolumn{9}{c}{\mbox{}}\\[-5ex]
\mbox{S}& \mbox{X}& & \mbox{AND}& \mbox{OR}& \mbox{ANDN}&
\mbox{ORN}& \mbox{XOR}& \mbox{EQU}\\

\hline
\multicolumn{9}{c}{\mbox{}}\\[-5ex]
\mbox{Low}& \mbox{Low}& & \mbox{Off}& \mbox{Off}& \mbox{Off}&
\mbox{On}& \mbox{Off}& \mbox{On}\\[-1ex]

\mbox{Low}& \mbox{High}& &  \mbox{Off}& \mbox{On}&
\mbox{Off}& \mbox{Off}& \mbox{On}& \mbox{Off}\\[-1ex]

\mbox{High}& \mbox{Low}& &  \mbox{Off}& \mbox{On}&
\mbox{On}& \mbox{On}& \mbox{On}& \mbox{Off}\\[-1ex]

\mbox{High}& \mbox{High}& &  \mbox{On}& \mbox{On}&
\mbox{Off}& \mbox{On}& \mbox{Off}& \mbox{On}\\
\hline
\multicolumn{9}{c}{\mbox{}}\\[-6ex]
\multicolumn{2}{c}{\mbox{Refs.}}& &
\mbox{\cite{Buchler03,PlosBiol06,Anderson07}}&
\mbox{\cite{Buchler03,PlosBiol06,Fernandez94}}&
\mbox{\cite{PlosComputBiol06}}& \mbox{\cite{Erill07}}&
\mbox{\cite{Buchler03}}&
\mbox{\cite{Buchler03,PlosComputBiol06}}\\
\hline\hline
\end{array}  \]
\end{table}
\begin{figure}
  \centering
  \includegraphics[width=12cm]{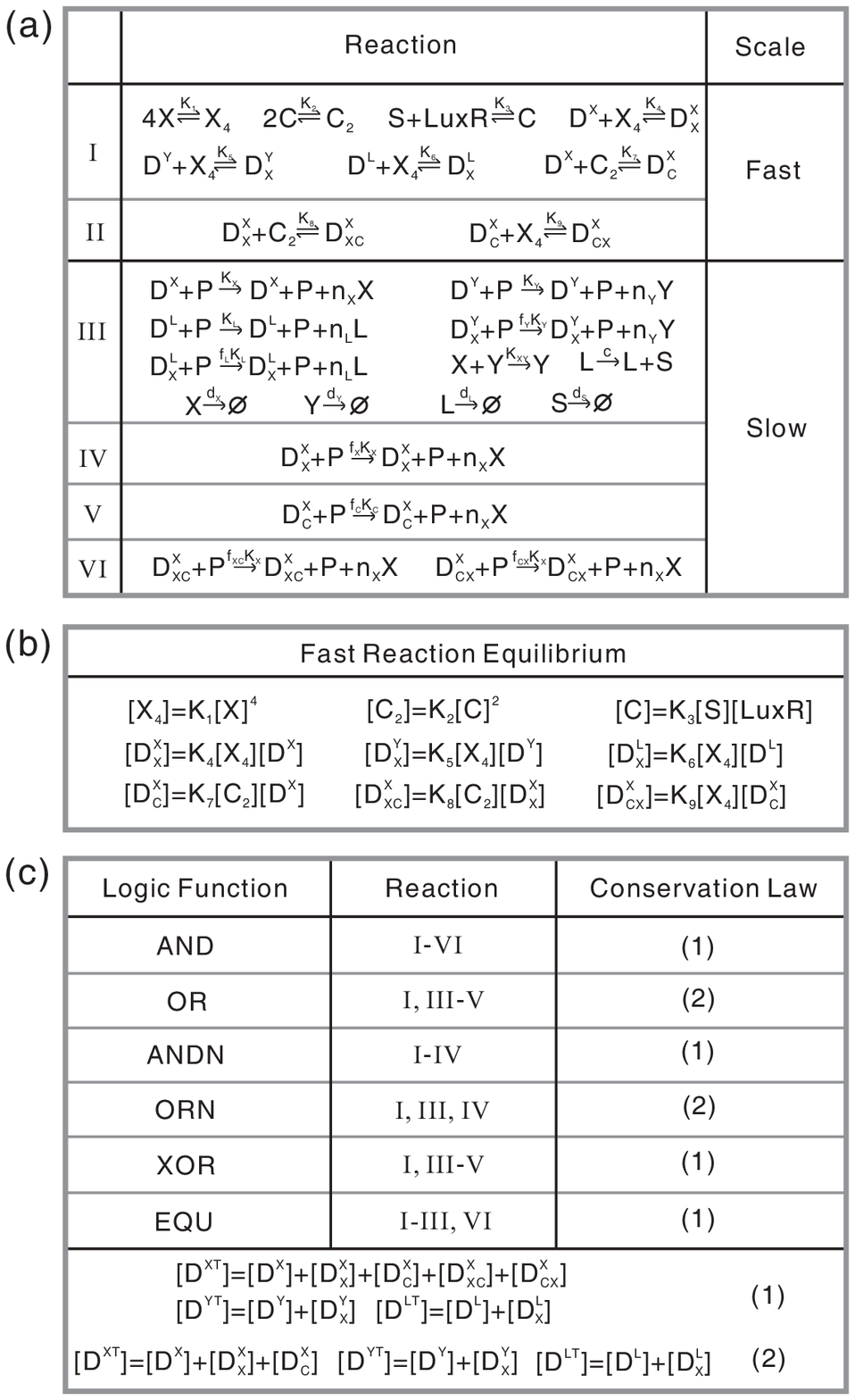}
  \caption{(a) The biochemical reactions are classified as two classes: fast and
  slow; (b) The equilibrium equations for the fast reactions;
  (c) Six logic operations and their biochemical reactions, where the corresponding conservation laws are
  listed on the bottom. The reaction rates used are experimentally reasonable, refer to \cite{DMcMillen-PNAS02}.}
  \label{fig2}
\end{figure}
Then, we define the chemical species in Table \textrm{II}. All
biochemical reactions are listed in Fig. \ref{fig2}(a), and some
reaction constants are listed in Table \textrm{III}. Assume the fast
reactions to be in equilibrium, refer to the equilibrium equations
shown in Fig. \ref{fig2}(b), where square brackets stand for
concentrations of species. In fact, the fast reaction equilibrium
trick based on quasi-steady state approximation approach has been
widely applied to reduce the complexity of multiscale problems
\cite{DMcMillen-PNAS02,Segel88}. The conservation laws for DNA
binding sites in the regulatory regions are listed on the bottom in
Fig. \ref{fig2}(c).
\begin{table}[htb]\label{t2}
\caption{Description of species in biochemical reactions}
\vspace{0.3cm}
\begin{tabular}{llll}
\hline\hline
Species & Descriptions & Species & Descriptions\\
\hline
X & protein CII & D$^{\rm X}$ & DNA binding site in cII gene\\

Y & protein FtsH & D$^{\rm Y}$ & DNA binding site in ftsH gene\\

L & protein LuxI & D$^{\rm L}$ & DNA binding site in luxI gene\\

S & autoinducer AHL & D${\rm _X^X}$ & CII-DNA complex\\

C & LuxR-S complex & D${\rm _C^X}$ & LuxR-S-DNA complex\\

X$_4$ & CII tetramer & D${\rm _{CX}^X}$ & CII-LuxR-S-DNA complex\\

C$_2$ & hetero-tetramer complex & D${\rm _{XC}^X}$ & CII-LuxR-S-DNA complex\\

P & RNA polymerase &  & \\
\hline\hline

\end{tabular}
\end{table}

\begin{table}[htb] \label{t3}
\caption{Descriptions and values of raw parameters.} \vspace{0.3cm}
\begin{tabular}{l|c|c}
\hline \hline Descriptions & Values & References\\\hline
Dimerization equilibrium constant & $K_1=1.8\times10^{18} {\rm M^{-1}}$, $K_2=K_3=2.5\times10^6 {\rm M^{-1}}$ & \cite{Ho, Ptashne02}\\
\hline
Regulatory binding constant & $K_4=K_5=K_6=K_7=5\times10^6 {\rm M^{-1}}$ & \cite{Ptashne02}\\
 & $K_8=K_9=2.5\times10^6 {\rm M^{-1}}$ & \\
\hline
Degradation rate of protein & $d_X=d_Y=\ln2/10{\rm min^{-1}}$,$d_L=\ln2/0.2{\rm min^{-1}}$ & \cite{note}\\
 & $d_S=\ln2/1.1{\rm min^{-1}}$ & \\
\hline
Autoinducer synthesis rate & $c=1.1{\rm min^{-1}}$ & \cite{More96}\\
\hline
Bulk rate of transcription and translation  & $k_X=k_Y=k_L=k_C=30{\rm min^{-1}}$ & \cite{Elowitz00,Ptashne02}\\
 \hline
Amplified factor of transcription rate  & $f_X=f_Y=f_L=10,f_C=90,f_{XC}=f_{CX}=90$ & \cite{Ptashne02}\\
\hline
The rate of repressor degradation by Y  & $K_{XY}=2\times10^{-5}{\rm M^{-1}}$ & \cite{DMcMillen-PNAS02}\\
\hline
Plasmid copy number   & $m_X=10, m_Y=1, m_L=50$ & \cite{Ptashne02}\\
\hline
Concentration of LuxR & $[LuxR]=1\times10^{-8}{\rm M^{-1}}$ & \cite{DMcMillen-PNAS02}\\
\hline
Other parameters & $n_XK_X[{\rm D^{XT}][{\rm P}]}=8\times10^{-8}{\rm M min^{-1}}, n_X=n_Y=n_L=1$ & \cite{Hasty-Chaos01}\\
\hline \hline
\end{tabular}
\end{table}

\begin{table}[htb] \label{t4}
\caption{Rescaled variables and rescaled parameters for models. We
assume
$K_2=K_3,\,K_4=K_5=K_6=K_7,\,K_{8}=K_{9},\,K_X=K_Y=K_L=K_C,\,[{\rm
D}^{\rm{XT}}]=[{\rm D}^{\rm{YT}}]=[{\rm D}^{\rm{LT}}]=[{\rm
D}^{\rm{ST}}],\,\, {\rm and}\,\,\,n_X=n_Y=n_L=n_S\triangleq e$ for
rescaling.} \vspace{-0.3cm} \vspace{-0.5cm}
\[ \begin{array}{ll}\hline\hline
\mbox{\hspace{0cm}Rescaled Variables} & \mbox{\hspace{0cm}Rescaled Parameters}\\
\hline

\D X\triangleq(K_4K_1)^{1/4}[{\rm X}]\hspace{3cm} & \mu_x \triangleq \D f_X, \mu_y \triangleq \D f_Y, \mu_l \triangleq \D f_L, \mu_s \triangleq \D f_C, \mu_{xs} \triangleq \D f_{XC}K_{8}/K_7+f_{CX}K_{9}/K_4\\[2ex]

\D Y\triangleq(K_4K_1)^{1/4}[{\rm Y}] & \alpha_x \triangleq \D m_X, \alpha_y \triangleq \D m_Yn_Y/n_X, \alpha_l \triangleq \D m_Ln_L/n_X, \tau^* \triangleq \D n_XK_X(K_1K_4)^{1/4}[{\rm D^{XT}][P]}\\[2ex]

\D L\triangleq(K_4K_1)^{1/4}[{\rm L}] & \alpha_s \triangleq \D cK_3(K_2K_7)^{1/2}[{\rm LuxR}]/((K_1K_4)^{1/4}t^*), t \triangleq \D \tau^*\tau,\delta_x \triangleq \D d_X/\tau^*, \delta_y \triangleq d_Y/\tau^*\\[2ex]

\D S\triangleq K_3(K_2K_7)^{1/2}[{\rm LuxR}][{\rm S}] & \delta_l \triangleq d_L/\tau^*, \delta_{xy} \triangleq \D K_{XY}/((K_1K_4)^{1/4}\tau^*), \delta_s \triangleq \D d_S/\tau^*, \lambda\triangleq \D K_{8}/K_7+K_{9}/K_4\\[2ex]
\hline\hline
\end{array}  \]
\end{table}

Define concentrations as our dynamical variables (see Table
\textrm{IV}). Using equalities for the fast reactions and the
conservation laws, we can eliminate fast variables. To that end, we
can derive expressions of five cis-regulatory input functions
(CRIFs) which are listed in Table \textrm{V}, and the rate equations
which describe the evolution of the concentrations of X, Y, L and S
monomers as follows
\begin{equation}\D
\begin{split}
\frac{dX_i}{dt}&={\rm CRIF}-\delta_{xy} X_iY_i-\delta_x X_i\\
\frac{dY_i}{dt}&=\alpha_y\frac{1+\mu_y X_i^4}{1+X_i^4}-\delta_y Y_i\\
\frac{dL_i}{dt}&=\alpha_l\frac{1+\mu_l X_i^4}{1+X_i^4}-\delta_l L_i\\
\frac{dS_i}{dt}&=\alpha_s L_i-\delta_s S_i+ \eta(S_e-S_i),
\end{split}
\end{equation}
where $S_e=\frac{Q}{N}\sum_{i=1}^NS_i$ (when $N$ cells are
considered) in which $Q$ depends on the cell density in a nonlinear
way. The rescaled parameters are also listed in Table \textrm{IV}.
\begin{table}[htb]\label{t5}
\caption{Biochemical reactions and cis-regulatory input functions
for relaxation oscillators.}\vspace{-0.3cm}
\[ \begin{array}{cccc}\hline\hline
\multicolumn{2}{c}{\mbox{Logic Function}} & \multicolumn{2}{c}{\mbox{\hspace{0.7cm}CRIF}} \\
\hline
\multicolumn{2}{c}{\rule[0cm]{0mm}{0.583cm}\mbox{AND}} & \multicolumn{2}{c}{\hspace{0.7cm}\D \alpha_x\frac{1+\mu_x X^4+\mu_s S^2+\mu_{xs} X^4S^2}{1 + X^4 + S^2 + \lambda X^4S^2}} \\[2ex]
\multicolumn{2}{c}{\mbox{OR}} & \multicolumn{2}{c}{\hspace{0.7cm}\D \alpha_x\frac{1+\mu_x X^4+\mu_s S^2}{1 + X^4 + S^2}} \\[2ex]
\multicolumn{2}{c}{\mbox{ANDN}} & \multicolumn{2}{c}{\hspace{0.7cm}\D \alpha_x\frac{1+\mu_x X^4}{1 + X^4 + S^2 + \lambda X^4S^2}} \\[2ex]
\multicolumn{2}{c}{\mbox{ORN}} & \multicolumn{2}{c}{\hspace{0.7cm}\D \alpha_x\frac{1+\mu_x X^4}{1 + X^4 + S^2}} \\[2ex]
\multicolumn{2}{c}{\mbox{XOR}} & \multicolumn{2}{c}{\hspace{0.7cm}\D \alpha_x\frac{1+\mu_x X^4+\mu_s S^2}{1 + X^4 + S^2 + \lambda X^4S^2}} \\[2ex]
\multicolumn{2}{c}{\mbox{EQU}} & \multicolumn{2}{c}{\hspace{0.7cm}\D \alpha_x\frac{1+\mu_{xs} X^4 S^2}{1 + X^4 + S^2 + \lambda X^4S^2}}\\
\hline\hline
\end{array}  \]
\end{table}

\subsection{Analysis}

\subsubsection{Phase reduction approach}

First, we rewrite the final equation of Eq. (1) as the following
symmetric form of coupling
\begin{eqnarray} \label{eq2}
\frac{dS_i}{dt}=\alpha_s L_i-\delta_s
S_i-\eta(1-Q)S_i+\frac{1}{N}\sum^{N}_{j=1}\eta Q(S_j-S_i)\,.
\end{eqnarray}
For convenience, the system composed of both the first three
equations of Eq. (1) and the equation
\begin{eqnarray} \label{eq3}
\frac{dS_i}{dt}=\alpha_s L_i-\delta_s S_i-\eta(1-Q)S_i
\end{eqnarray}
is called as auxiliary system, which is assumed to generate a
sustained oscillation. Then, we perform an analytical study of the
entire system in the phase model description, which holds in a weak
coupling case \cite{BookKuramoto84}. The main steps are as follows.
For convenience, we express the system of globally coupled
oscillators as
\begin{eqnarray} \label{eq31}
\frac{dx_i}{dt}=
f(x_i)+\frac{1}{N}\sum^{N}_{j=1}p(x_i,\,x_j),\,1\leq i\leq N,
\end{eqnarray}
where $x_i=(X_i,\,Y_i,\,L_i,\,S_i)^{\rm T}$, $
f=(F_1,\,F_2,\,F_3,\,F_4)^{\rm T}$ with $F_1={\rm CRIF}-\delta_{xy}
X_iY_i-\delta_x X_i$, $F_2= \alpha_y\frac{1+\mu_y
X_i^4}{1+X_i^4}-\delta_y Y_i$, $F_3= \alpha_l\frac{1+\mu_l
X_i^4}{1+X_i^4}-\delta_l L_i$ and $F_4=\alpha_s L_i-\delta_s
S_i-\eta(1-Q)S_i$, and $p(x_i,\,x_j)=(0,0,0,\eta Q(S_j-S_i))^{\rm
T}$. Assume that the uncoupled oscillator has period $T$. By
Kuramoto's theorem, for a weakly perturbed system we can obtain the
corresponding phase model:
\begin{eqnarray} \label{eq32}
\frac{d\phi_i}{dt}=1+\frac{1}{N}Z(\phi_i)\cdot\sum^{N}_{j=1}
p(x_i(\phi_i),\,x_j(\phi_j)),
\end{eqnarray}
where each $x_i(\phi_i)$ is the point on the limit cycle having
phase $\phi_i$, the symbol `$\cdot$' is the dot product of two
vectors, and
\begin{eqnarray} \label{eq33}
Z(\phi_i)={\rm grad} \,\phi_i(x_i).
\end{eqnarray}
$Z(\phi_i)$, a phase response function characterizing the phase
advance per unit perturbation, is a $2\pi$-period function with
respect to $\phi_i$. To study collective properties of the network,
such as synchronization and clustering, it is convenient to
represent each $\phi_i$ as $\phi_i=t+\vartheta_i$ with the first
term capturing the fast free-running natural oscillation
$d\phi_i/dt=1$, and the second term capturing the slow
network-induced build-up of phase derivation from the natural
oscillation. Substituting the expression of $\vartheta_i$ into Eq.
(\ref{eq32}) results in
\begin{eqnarray} \label{eq33}
\frac{d\vartheta_i}{dt}=\frac{1}{N}Z_i(t+\vartheta_i)\cdot\sum^{N}_{j=1}
p(x_i(t+\vartheta_i),\,x_j(t+\vartheta_i)),
\end{eqnarray}
The classical method of averaging consists in a near-identity change
of variables that transforms the system into the form
\begin{eqnarray} \label{eq4}
\frac{d\phi_i}{dt}=
1+\frac{1}{N}\sum^{N}_{j=1}H_{ij}(\phi_j-\phi_i)\,,
\end{eqnarray}
where $H_{ij}(\Delta \phi)$ represents the interaction function with
respect to the phase difference $\Delta \phi=\phi_j-\phi_i$ between
two cells,
\begin{eqnarray} \label{eq5}
H_{ij}(\phi_j-\phi_i)=\frac{1}{T}\int_{0}^{T}Z_i(t)\cdot
p(x_i(t),\,x_j(t+\phi_j-\phi_i))dt
=\frac{1}{2\pi}\int_{0}^{2\pi}Z_i(\theta)\cdot
p(\phi_j-\phi_i+\theta)d\theta
\end{eqnarray}
which can be calculated numerically \cite{Ermentrout91}. In what
follows, we omit subscripts $i$ and $j$ for convenience. From
$H(\Delta \phi)$, we introduce a function: $G(\Delta \phi)= H(\Delta
\phi)-H(-\Delta \phi)$, to determine the mode of coupling. If
$G(\Delta \phi)$ exhibits a positive slope at $\Delta \phi=0$, i.e.,
$G'(0)>0$, the coupling is phase-attractive; If $G'(0)<0$, the
coupling is phase-repulsive. Such an approach based on the sign of
$G'(0)$ that depends generally on the intrinsic dynamics of the
uncoupled oscillator and on the interaction between the oscillators
is more effective than that of directly observing the network
topology in determining the mode of weak coupling \cite{Ullner07},
especially in the case of complex network architectures.

According to Tables \textrm{III} and Table \textrm{IV}, we can
estimate our system parameter values as follows: $\alpha_x=10$,
$\alpha_y=1$, $\alpha_l=50$, $\alpha_s=0.4$, $\delta_x=0.5$,
$\delta_y=0.5$, $\delta_l=25$, $\delta_s=45$, $\delta_{xy}=5$,
$\mu_x=10$, $\mu_y=10$, $\mu_l=10$, $\mu_s=9$, $\mu_{xs}=90$,
$\lambda=1$, $\eta=10$, $Q=0.5$. For such a set of values, numerical
simulation verifies that the term $\frac{1}{N}\sum^{N}_{j=1}\eta
Q(S_j-S_i)$ affects the timing but not the amplitude of the
auxiliary system for any $N\geq2$, so the above analysis is
feasible. In addition, we emphasize that for other different
experiments on multicellular systems with the quorum sensing
\cite{You,Balagadde05,
Brenner,Basu04,Basu05,Halsintine,Balagadde08}, the differences
between the rescaled parameter values are not so large that they
abolish our conclusions.

\subsubsection{Determining the stability of balanced clustering}

Balanced clusters mean that $N$ oscillators are divided into $M$
subgroups of the equal cell number with each subgroup being
synchronized and with the equal phase difference between neighboring
subgroups. Here, we employ Okuda's approach \cite{Okuda93} to
determine the stability of such clusters (see the Appendix of this
paper for details). In that method, we need to calculate two kinds
of eigenvalues: one is associated with intra-cluster fluctuations
and the other with inter-cluster fluctuations, which are denoted by
$\lambda_p$ and $\lambda_q$ respectively, where $M\leq p\leq N-1$
and $0\leq q\leq M-1$ with $M$ being the number of clusters
presumptively. For convenience, denote by $\lambda^{(1)}$ and
$\lambda^{(2)}$ the $N-M$ same eigenvalues $\lambda_p$ and the
maximum of the the real parts of ($M-1$) non-zero eigenvalues
$\lambda_q$, respectively. By calculation, we find
$\lambda_p=1/M\sum^{M-1}_{k=0}\Gamma'(2\pi k/M),p=M,M+1,\ldots,N-1$,
and $\lambda_q=1/M\sum^{M-1}_{k=0}\Gamma'(2\pi k/M)(1-\exp(-i2\pi
kq/M)),q=0,1,\ldots,M-1$, where $\Gamma(\Delta\phi)=H(-\Delta\phi)$.
Then, the stability of clusterings can be determined by the signs of
$\lambda^{(1)}$ and $\lambda^{(2)}$. Specifically, the clustering is
stable if both $\lambda^{(1)}$ and $\lambda^{(2)}$ are negative, and
unstable if $\lambda^{(2)}$ is positive. In addition, if
$\lambda^{(1)}$ is positive and $\lambda^{(2)}$ is negative, and
further if $M=N$, the $M$-cluster (i.e., the splay state) are also
stable.

In the Secs. \textrm{III} and \textrm{IV}, we will numerically study
cooperative behaviors of coupled genetic relaxation oscillators with
different CRMs. In contrast to the previous works
\cite{DMcMillen-PNAS02,JGarciaOjalvo-PNAS04,AKuznetsov-SIAM04,
TSZhou-PRL05,TSZhou-PlosOne07,Ullner07,Koseska_PRE07_1,
Koseska_PRE07_2,Koseska_PRE07_3,Yuan_PhysRevE08,chaos-08,Ullner_PRE08},
we will show that different CRMs can drive fundamentally different
dynamic patterns.

\section{Case of two coupled cells: phase locking}

Synchronization and clustering of genetic oscillators coupled to
quorum sensing result from the interplay between the intrinsic
properties of the individual cells, the type of cellular
communication, and the network topology. To gain insights into the
rules governing dynamic patterns in complex networks of cells, here
we investigate the case of two coupled genetic oscillators in
detail.

\begin{figure}
  \centering
  \includegraphics[width=12cm]{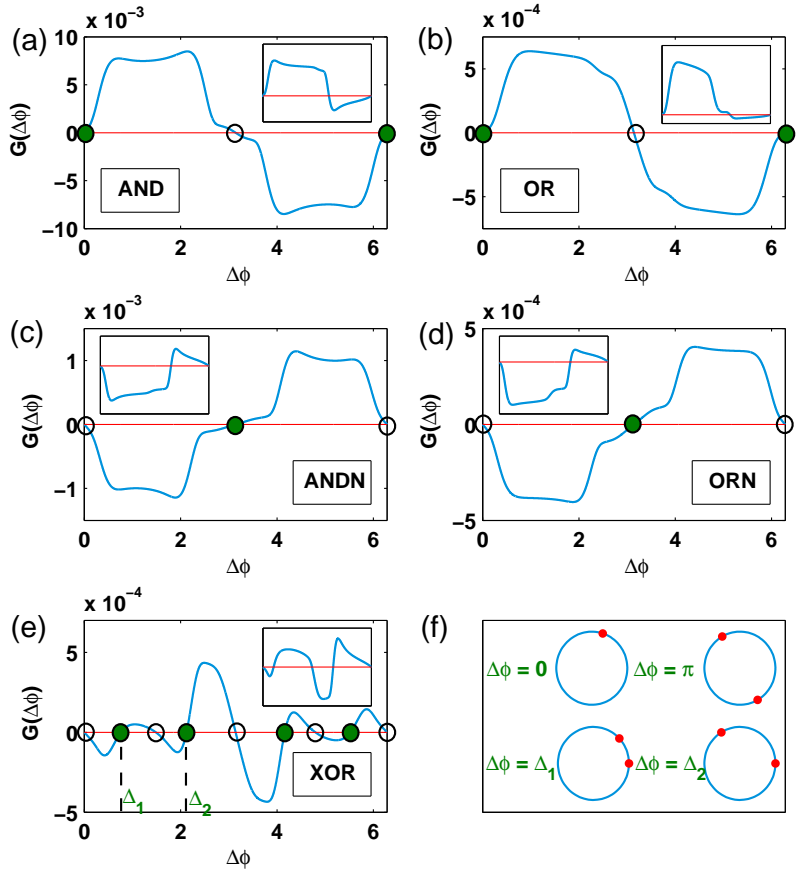}
  \caption{(Color online) The dependence of the function $G$ on
  phase difference $\triangle\phi$ in $[0,\,2\pi]$ and the distribution of its zero points in the case of two coupled oscillators.
  In (a)-(d), $G$ has three zero points, two of which are unstable and the other is one unstable in the cases of AND and OR
  whereas one is stable and the other two are unstable in the cases of ANDN and ORN.
  In (e), $G$ corresponding to XOR has 4 stable zero points and five unstable zero
  points. In all the cases, filled circles represent stable points and open circles do unstable
  points, and insets display the corresponding $H(\triangle\phi)$ of the auxiliary system. In (f), some typical instantaneous distributions
  of phases are demonstrated in the five cases of logic operations.}
  \label{fig3}
\end{figure}

First, based on Eq. (\ref{eq4}) the phase model of two coupled
oscillators can be characterized by
\begin{equation}\D \label{eq6}
\begin{split}
\frac{d\phi_1}{dt}&=1 + H(\Delta\phi)\\
\frac{d\phi_2}{dt}&=1 + H(-\Delta\phi),
\end{split}
\end{equation}
where the phase difference is denoted as $\Delta\phi=\phi_2-\phi_1$.
The interplay between the two oscillators is often described by the
evolution of the phase difference $\Delta\phi$, which is determined
solely by the odd part of the effective coupling function
$G(\triangle\phi)$, i.e., $H(\triangle\phi)-H(-\triangle\phi)$. That
is, the dynamics of $\Delta\phi$ is given by
\begin{eqnarray} \label{eq7}
\frac{d\Delta\phi}{dt}&=-G(\triangle\phi)
\end{eqnarray}
The zero points of $G(\triangle\phi)$ are the fixed points of Eq.
(\ref{eq7}). These fixed points describe the phase-locked states of
two coupled cells and their stabilities are determined by the sign
of slope of the curve $G(\triangle\phi)$ at the zero points: A
positive slope means that the corresponding fixed point is stable,
implying that $\triangle\phi$ nearby the fixed point dynamically
converges to the fixed point, whereas a negative slope means that
the fixed point is unstable, implying that $\triangle\phi$ close to
the fixed point dynamically diverges. The size of the slope
determines the convergence or divergence rate at the fixed point.
The function $G(\triangle\phi)$ corresponding to five logic
operations AND, OR, ANDN, ORN or XOR is shown in Fig.
\ref{fig3}(a)-(e) respectively (here and below we did not
investigate the case of EQU due to the fact that the EQU destroys
the dynamics of the core relaxation oscillator in the region of
biological reasonable parameters, leading to the loss of sustained
oscillation), whereas the interaction function $H(\triangle\phi)$s
is shown in the insets.

(1) AND and OR. The function $G(\triangle\phi)$ in Figs.
\ref{fig3}(a)-(b) equates to zero at $\triangle\phi=0$ with positive
slope and at $\triangle\phi=\pi$ with negative slope. Moreover, the
zero point $\triangle\phi=0$ is the unique stable state of Eq.
(\ref{eq7}). Therefore, the phase-model analysis predicts that the
phase difference of any initial values except $\triangle\phi=\pi$
eventually converges to $\triangle\phi=0$. This result is also
verified by integrating the original model with various initial
values. A typical snapshot is plotted in the upper-left of Fig.
\ref{fig3}(f). Thus, the analysis together with numerical simulation
shows that AND and OR play a role in stabilizing the in-phase
synchronization for two coupled cells. In this case, the coupling is
phase-attractive.

(2) ANDN and ORN. Equation (\ref{eq7}) has one unstable state
$\triangle\phi=0$ and one stable state $\triangle\phi=\pi$, both of
which correspond to zero points of the function $G$, as shown in
Fig. \ref{fig3}(c)-(d). The role of ANDN and ORN is to stabilize the
antiphase state and prevent the in-phase state. More precisely, the
integration between the intracellular activator and intercellular
signaling repressor in our model destabilizes the in-phase
synchronization. In this case, the coupling is phase-repulsive. A
typical snapshot is plotted in the upper-right of Fig.
\ref{fig3}(f).

(3) XOR. The function $G$ has nine zero points, four of which,
denoted by $\triangle\phi=\triangle_1,\,\triangle_2$ and
$\triangle\phi=\pi+\triangle_1,\,\pi+\triangle_2$, respond to stable
states of Eq. (\ref{eq7}) and the other five to unstable states, as
shown in Fig. \ref{fig3}(e). The unstable states form the boundaries
for the attraction basins of the stable states. The role of XOR is
to stabilize four out-of-phase states with phase difference
$\triangle\phi=\triangle_1,\,\triangle_2$ and
$\triangle\phi=\pi+\triangle_1,\,\pi+\triangle_2$ respectively,
whereas to destabilize the in-phase and anti-phase state. Two
typical snapshots are shown in the bottom row of Fig. \ref{fig3}(f).

\section{Case of a population of cells: synchronization and clustering}

In this section, we investigate the case of $N$ coupled genetic
oscillators ($N>2$), focusing on two dynamical behaviors, i.e.,
synchronization and clustering, which are ensemble phenomena
observed commonly in natural and artificial populations of (possibly
weakly) interacting oscillators. Synchronization is a cooperative
in-phase behavior, which has been the subject of numerous studies in
physics and biology
\cite{BookKuramoto84,APikovskyBook01,StrogatzBook03,ManrubiaBook04},
whereas clustering is a fragmentation of the collective behavior in
locally synchronized but well separated subgroups, which has been
also observed in numerous contexts with distinct contributions
\cite{Golomb92,Okuda93,Kiss05,Kiss07,AFTaylor08,Reynolds_ComputGraph87,
Pogromsky_PhysicaD02,Belykh_Chaos03,Golubitsky_SIAM05,Golubitsky_BullAmerMathSoc06}.
In what follows, we investigate balanced clustering and non-balanced
clustering separately for clarity (note that synchronization is a
particular type of clustering, i.e., 1-cluster).

\subsection{Balanced clustering}

In the analysis part, we present an approach for determining the
stability of balanced clustering. Here, we display numerical results
for balanced clustering. In particular, we show that CRMs of the
different structure play different roles in the achieving of
collective behaviors.

\begin{figure}
  \centering
  \includegraphics[width=12cm]{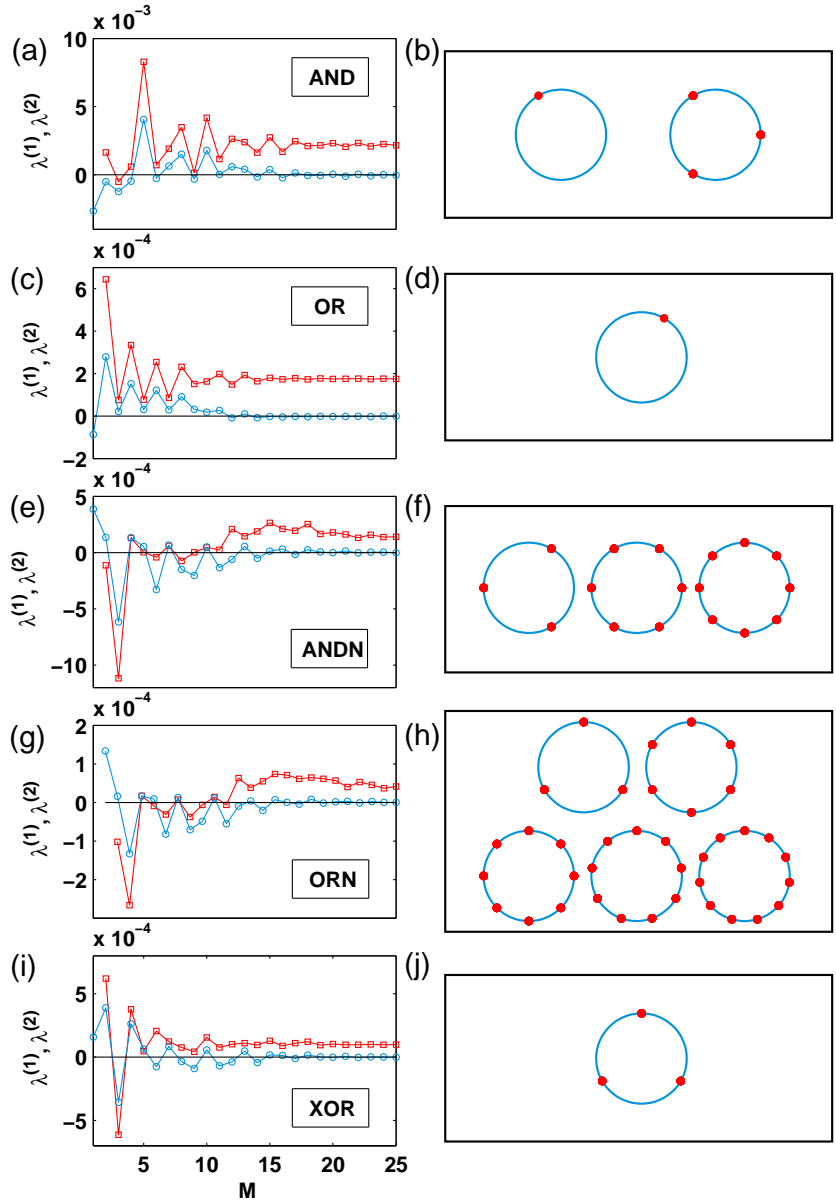}
  \caption{(Color online) (Left panel) Eigenvalues associated with
  intra-cluster fluctuations $\lambda_{1}$ (blue cycle) and the
  maximal real  part of non-zero eigenvalues associated with inter-cluster
  fluctuations $\lambda_{2}$ (red square) as a function of the number
  of balanced clusters with five different logic functions; (Right panel)
  The corresponding instantaneous phase distribution
  of all possible balanced clusters: (a)-(b) 1- and 3-cluster states
  for AND; (c)-(d) 1-cluster state (complete synchronization) for OR;
  (e)-(f) 3-, 6- and 8-cluster states for ANDN; (h)-(i) 3, 6-, 8-, 9-
  and 11-cluster states for ORN; (j)-(k) 3-cluster state for XOR. Different clustering states
  appear due to different choices of initial conditions but the cell number is fixed as $N=792$.}
  \label{fig4}
\end{figure}

\begin{figure}
  \centering
  \includegraphics[width=12cm]{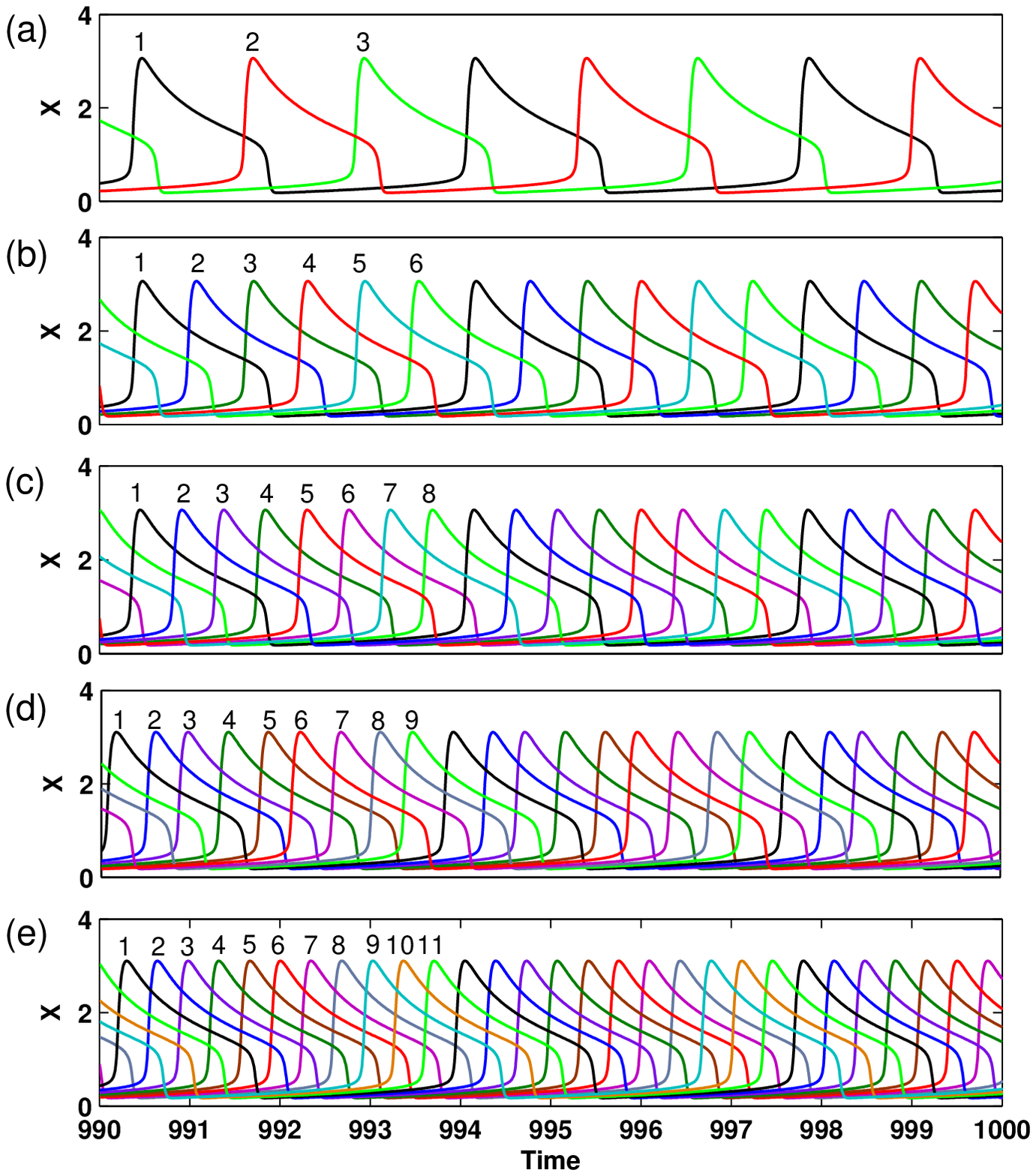}
  \caption{(Color online) Temporal evolutions of the concentration
  of X corresponding to ORN in Fig. \ref{fig4}: (a) 3-cluster state; (b) 6-cluster state;
  (c) 8-cluster state; (d) 9-cluster state; (e) 11-cluster state, where each cluster state is
  indicated by different color or an integer. For each obtained cluster state, numerical integration begins from an initial condition
  close to the corresponding clustering, and plot shown begins after allowing a transient time
  of $10^4$ units.}
  \label{fig5}
\end{figure}

(1) AND. Figure \ref{fig4}(a) indicates that $1$- and $3$-cluster
states are stable since both $\lambda_1$ and $\lambda_2$ are
negative. The instantaneous phase distributions on the unit cycle as
shown in Fig. \ref{fig4}(b) verify the coexistence of stable
complete synchronization and $3$-cluster state.

(2) OR. In this case, the eigenvalues shown in Fig. \ref{fig4}(c)
indicate that only the complete synchronization ($1$-cluster) is
stable, which is verified by the numerical simulation shown in Fig.
\ref{fig4}(d). The analysis together with the numerical simulation
shows that the OR plays a role of stabilizing complete
synchronization, i.e., for any initial conditions for these
oscillators, the systems consequentially evolve into a stable
complete synchronization.

(3) ANDN. The stability analysis of the eigenvalues shown in Fig.
\ref{fig4}(e) reveals that the network of coupled oscillators with
the ANDN possesses complex cluster-balanced states, e.g., the stable
$3$-, $5$- and $8$-cluster states. These clustering states are
numerically implemented as shown in Fig. \ref{fig4}(f).

(4) ORN. We give the results on the stability analysis as shown in
Fig. \ref{fig4}(g), which indicate that the population of
oscillators can give rise to more complex cluster-balanced states
than those displayed in the case of ANDN, e.g., two additional
cluster-balanced states, $9$- and $11$-cluster states are found. The
instantaneous phase distributions of these clustering states on the
unit cycle are shown in Fig. \ref{fig4}(h).

(5) XOR. In this case, the system of coupled oscillators possess
only a stable cluster-balanced state ($3$-cluster) that can be seen
from the sign of two eigenvalues determining the stability (see Fig.
\ref{fig4}(i)). A snapshot of the unique balanced clustering is
shown in the Fig. \ref{fig4}(j).

To display cellular patterns more clearly, we also plot all the time
courses of the component X in the case of ORN, refer to Fig.
\ref{fig5}. These cluster states appearing in the cases of different
logic operations indicates that different CRMs can drive
fundamentally different cellular patterns.

\subsection{Non-balanced clustering}

\begin{figure}
  \centering
  \includegraphics[width=12cm]{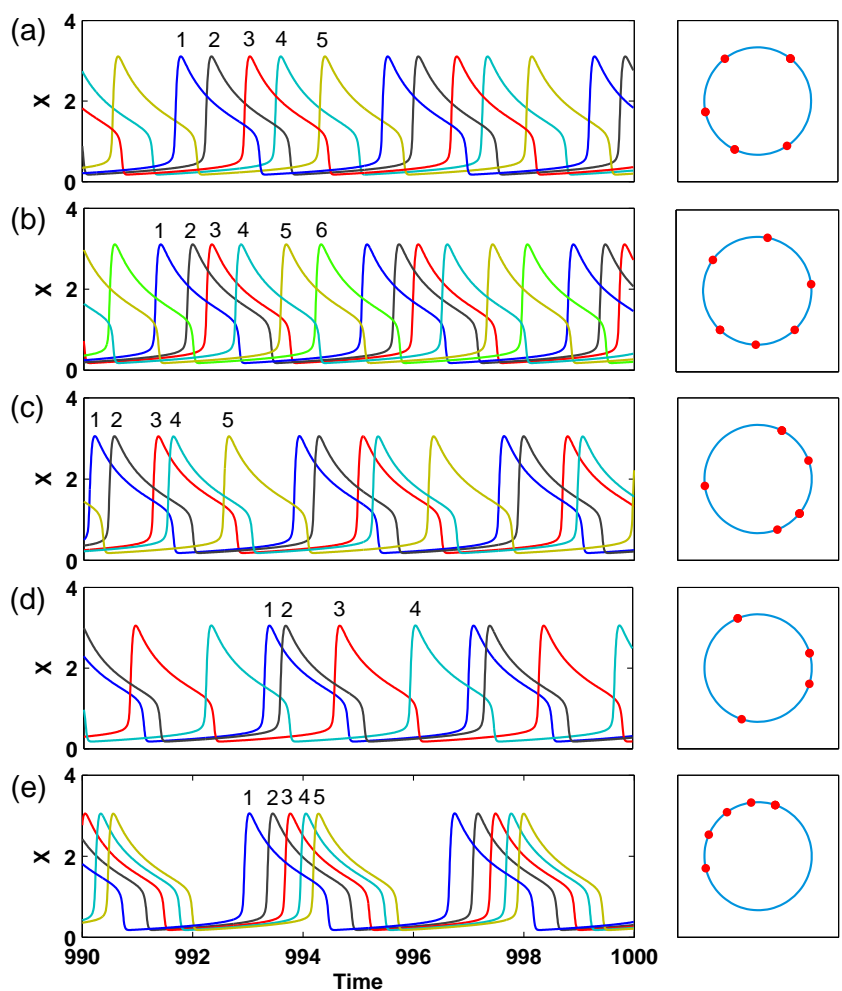}
  \caption{(Color online) Non-balanced clusterings found in the cases of ORN and XOR. (left panel) Temporal evolutions of the concentration
  of X corresponding to ORN and XOR and (right panel) cellular patterns: (a) 5-cluster state for ORN; (b) 6-cluster state for ORN;
  (c) 5-cluster state for XOR; (d) 4-cluster state for XOR; (e) 5-cluster state for XOR, where each cluster state is
  indicated by different color or an integer. For each shown non-balanced clustering, numerical integration begins from an initial condition
  close to the corresponding clustering, and plot shown begins after allowing a transient time
  of $10^4$ units.}
  \label{fig6}
\end{figure}

Except for balanced clustering as shown in the previous subsection,
we also find non-balanced clustering. However, finding all
non-balanced clusterings is much more difficult than finding all
balanced clusterings since the former, one needs to search for all
stable regions of initial values of coupled systems that lead to
stable non-balanced clusterings, and this is even impossible only
with computer simulation when the cell number is large. Here, we
mainly want to show that non-balanced clusterings are existent in
some cases of five logic operations. By numerical simulation, we
find that different CRMs can drive different types of non-balanced
clusterings except for the OR case (since the complete
synchronization is globally stable). Taking the cases of ORN and XOR
as examples, we find several typical non-balanced clusterings which
are displayed in Fig. \ref{fig6}: In the case of ORN, we find 5- and
6-cluster states whereas in the case of XOR, we find a 4-cluster
state and two different types of 5-cluster states. Note that we did
not search out all non-balanced clusterings, and other types of
non-balanced clusterings except for those found are possible, but
would depend on the number of oscillators and the choice of initial
conditions.

\section{Effect of rewiring network on synchronization and clustering}

Biological rhythm results from the interplay between the intrinsic
properties of the individual cells, the properties of the
communication, as well as the network topology. Each property may
play an important role in shaping the emergent synchronous behavior.
Except that different CRMs can drive different cellular patterns
shown in the above two sections, the rewiring architecture of
individual cells also may play a significant role in promoting
synchronization or antisynchronization of coupled cells, e.g., it
has been shown that rewired interaction in a repressilator
population with cell-cell communication can offer diverse dynamics,
such as multistability and clustering \cite{chaos-08,Ullner07}. Note
that in the investigated-above models, the signal molecule regulates
an activator, thus performing an activator-regulated communication.
Due to biological background of the core genetic oscillator and the
quorum sensing, however, the signal molecule can also regulates a
repressor, leading to so-called repressor-regulated communication in
contrast to activator-regulated communication. In this section, we
investigate the effect of this rewiring architecture of motifs on
synchronization and clustering.

\begin{figure}
  \centering
  \includegraphics[width=10cm]{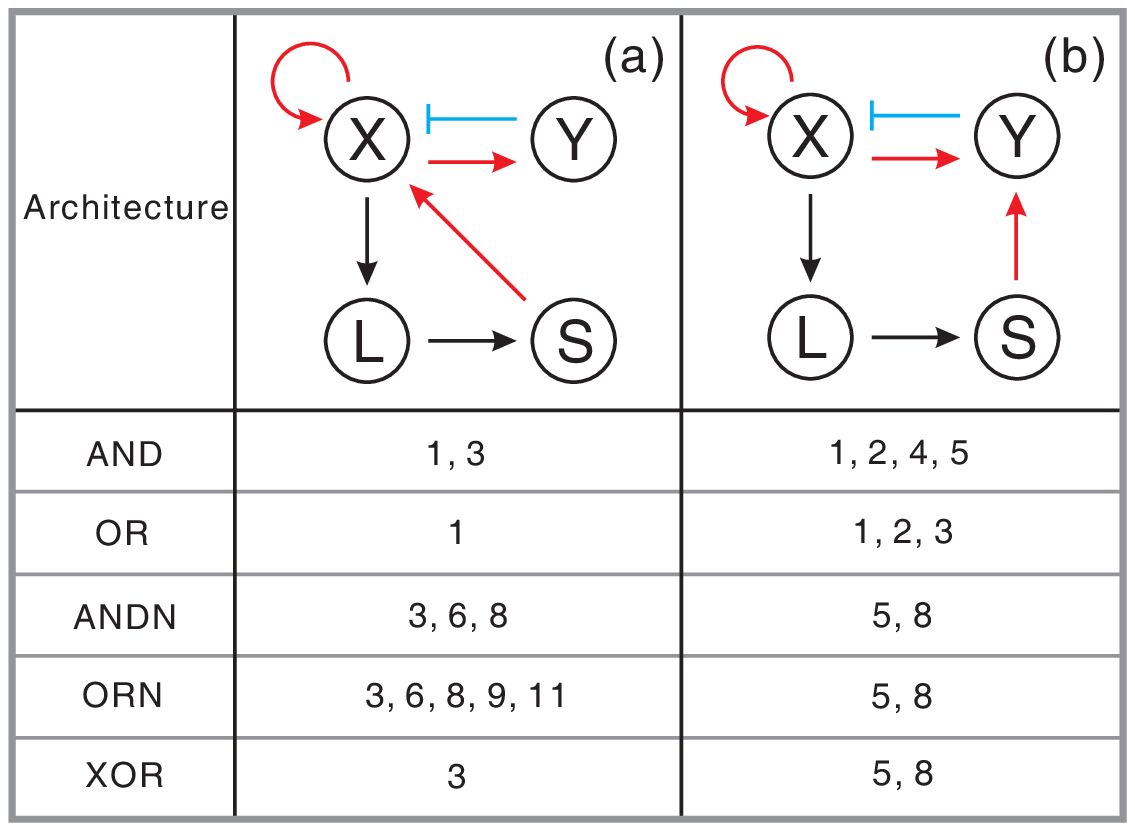}
  \caption{(Color online) Different balanced clusterings of original and rewired  genetic architectures.
  (a) activator-regulated communication; (b) repressor-regulated communication.
  The corresponding clusterings for two cases are listed on the bottom, respectively. Note: Only the same balanced clusterings
  are shown for three logic operations in the case of repressor-regulated communication, but different non-balanced clusterings
  are possible (data are not shown here).}
  \label{fig7}
\end{figure}

\begin{figure}
  \centering
  \includegraphics[width=12cm]{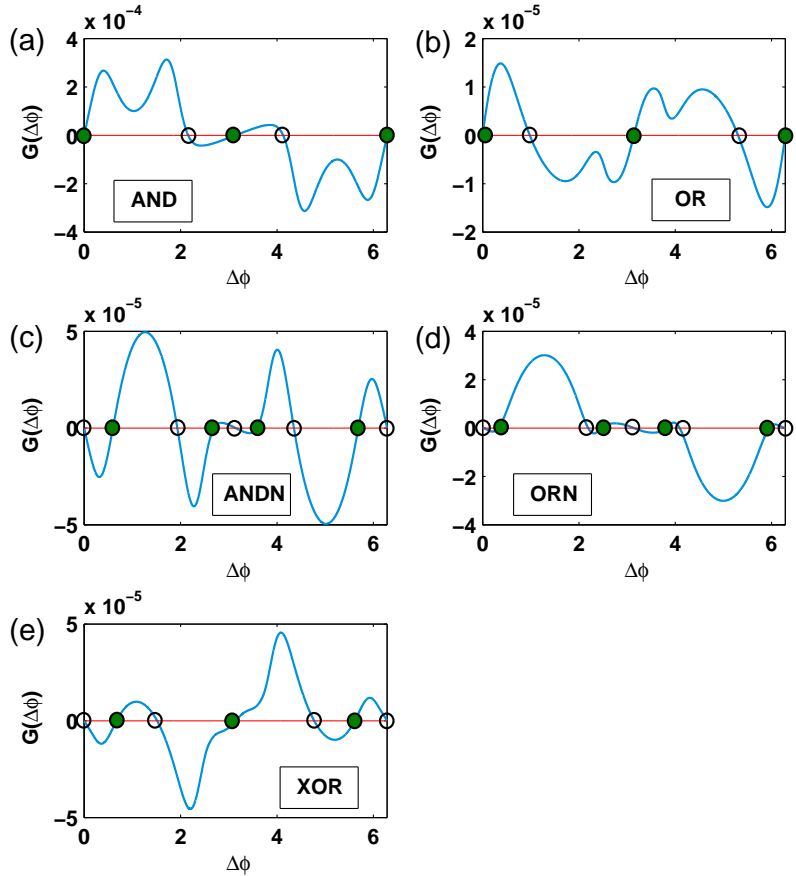}
  \caption{(Color online) The dependence of the function $G(\triangle\phi)$ on phase difference $G(\triangle\phi)$,
  where its stable zero points (filled circle) and unstable zero points
  (open circle) are shown.}
  \label{fig8}
\end{figure}

In contrast to the scheme of signal integration in the previous two
sections (refer to a simplified scheme shown in Fig. \ref{fig7}(a)),
in what follows we rewire the interaction of the signaling molecule
and its regulated gene inside the cell
\cite{Yuan_PhysRevE08,Halsintine}, as shown in Fig. \ref{fig7}(b).
More precisely, we let the signaling molecule AI and the TF X
combinatorially regulate the target gene y instead of gene x.
Completely similarly, we can derive expressions of 6 possible CRIFs
(see Table \textrm{V}), and the dynamical equations describing the
time evolution of the concentrations of X, Y, L and S monomers in
the following form:
\begin{equation}\D
\begin{split}
\frac{dX_i}{dt}&=\alpha_x\frac{1+\mu_x X_i^4}{1+X_i^4}-\delta_{xy} X_iY_i-\delta_x X_i\\
\frac{dY_i}{dt}&={\rm CRIF}-\delta_y Y_i\\
\frac{dL_i}{dt}&=\alpha_l\frac{1+\mu_l X_i^4}{1+X_i^4}-\delta_l L_i\\
\frac{dS_i}{dt}&=\alpha_s L_i-\delta_s S_i+ \eta(S_e-S_i),
\end{split}
\end{equation}
where the CRIFs are similar to those in the case of
activator-regulated communication, refer to Table \textrm{V} except
that parameters $\alpha_x$ and $\mu_{xs}$ are replaced by $\alpha_y$
and $\mu_{ys}$, respectively. In both cases, the settings of
parameter values are also the same except for $\mu_{ys}=90$. The
numerical results are summarized in Fig. \ref{fig7}, where all
balanced clusterings are listed in two cases of activator-regulated
communication and repressor-regulated communication for comparison.
From Fig. \ref{fig7}, we see that different CRMs also can drive
fundamentally different cellular patterns in the case of
repressor-regulated communication, but the wave patterns are
different from those in the case of activator-regulated
communication (Data for comparison are not shown). In addition, we
show how the odd part of the interaction function
$H(\triangle\phi)$, $G(\triangle\phi)$, in the five logic
operations, changes with phase difference
$\triangle\phi\in[0,\,2\pi]$ in Fig. \ref{fig8}, where stable zero
points (symbolled by filled circle) and unstable zero points
(symbolled by open circle) are shown. Our results suggest that the
architecture of biological systems might make them particularly
evolvable, namely, simple shuffling of finely-tuned network
architectures may render new functionalities of networks with
feedforward and feedback.

The rational design of biological networks and pathways promises to
reveal ways of rewiring cells for new biological functions or of
gaining insights into the behavior of natural systems. Much of the
work to date has focused on the manipulation of transcriptional and
post-transcriptional elements to create synthetic gene networks with
desired functions
\cite{Bayer05,Becskei00,Elowitz00,Gardner_Nature00,RGuantes06}. In
contrast, our present study provides a possible arsenal for
designing and constructing a network of genetic oscillators with
different cellular behavior, indicating that rationally
reprogramming integration of two input TFs by changing a CRM to
activate a targeted gene could be used to induce transition among
various cellular patterns towards the corresponding desired
functions. In spite of this, we expect that understanding how
different CRMs render different responses for the coupled genetic
oscillators with quorum sensing would provide a valuable insight
into designing new synthetic genetic circuits.

\section{Conclusion and discussion}

Using models of synthetic genetic relaxation oscillators coupled by
quorum sensing, we have shown both analytically and numerically that
different CRMs drive fundamentally different cellular patterns, such
as synchronization and balanced clustering, and non-balanced
clustering, by considering two types of communications:
activator-regulated communication and repressor-regulated
communication. Specifically, in the case of two coupled oscillators,
we have shown that different CRMs have marked influences on
characteristics of phase-locking processes, e.g., two oscillators
can display in-phase, anti-phase and out-of-phase synchronization
with a certain constant phase difference, depending on the type of
CRM. In the case of $N(>2)$ coupled oscillators with
activator-regulated communication, there are 1- and 3-balanced
clusters for AND, only 1-balanced cluster for OR, 3-, 6- and
8-balanced clusters for ANDN, 3-, 6-, 8-, 9- and 11-balanced
clusters for ORN, and only 3-balanced cluster for XOR, whereas in
the case of $N$ coupled cells with repressor-regulated
communication, there are 1-, 2-, 4- and 5-balanced clusters for AND,
1-, 2- and 3-balanced cluster for OR, 5- and 8-balanced clusters for
ANDN, ORN and XOR. In addition, some non-balanced clusters have been
also found. These results would provide a strategy for a network of
genetic oscillators: the selection of cooperative rhythmic manner,
e.g. synchronization and clustering, is governed by the nature of
the integration of the intracellular signal and the secretion of the
biochemical signals through which the oscillating cells are globally
coupled. In particular, genetic network architecture found in
synchronous circadian clocks \cite{JCDunlap-Cell99} might be
constrained since the complete synchronization independent of
initial conditions takes place only in the case of OR type of
response. In addition, our results would imply that multicellular
organisms evolve into some functional CRMs for particular goals
(e.g., cellular patterns) by performing an elaborate computation for
input TFs.

We expect that our findings will stimulate further investigations
under a more realistic condition involving stochasticity
\cite{TSZhou-PRL05,JMRaser05,NBarkai-Nature00,Zhang-PRE08} and
heterogeneity \cite{JGarciaOjalvo-PNAS04} as specified in the
following four points:

(1) In a stochastic environment, we should consider the stability of
the obtained desired dynamic pattern. Theoretically, Golomb
\textit{et al.}, have shown that the clustering state is stable on
the condition that noise intensity is below a critical value
\cite{Golomb92}. On the other hand, the global noise can enhance the
extent of phase synchronization \cite{Kawamura08}, but also can
destroy the clustering state like in slow switching \cite{Kori01}.
Therefore, we should carefully design the CRMs structure in the
presence of noise to preserve the desired dynamic patterns.

(2) In our model, a population of identical oscillators communicate
with a uniform coupling, but it would be of great interest to study
the influence of the cellular variability and coupling strength
heterogeneity on the synchronization and clustering. If
heterogeneity is sufficiently small compared to the coupling
strength, we can treat the system as identical oscillators, and
otherwise, the effect of heterogeneity should be considered. In
fact, it has been shown that heterogeneous coupling strength and
element variability can make the occurrence of clustering states
possible in networks of neural oscillators \cite{Li03}. Similarly,
in our case, heterogeneity would result in synchronization and
clustering.

(3) Our results were obtained under the condition that the
intercellular communication is rather weak. However, it is likely
that coupling is stronger than that considered here
\cite{Ullner07,Golomb94}. Therefore, it would be of interest to
analyze dynamical patterns in the case of strong coupling. In this
case, other modes of complex behaviors such as multistability
\cite{Ullner07,Koseska_PRE07_1}, inhomogeneous limit cycle
\cite{Ullner07,Ullner_PRE08}, oscillation death
\cite{AKuznetsov-SIAM04,Ullner07}, aperiodic oscillation
\cite{Gonze08}, and chaos \cite{Ullner_PRE08,Gonze08} may also
appear in our models.

(4) We point out that our results are in general robust to changes
in parameter values if they are not chosen close to the margin of
oscillation of the uncoupled oscillator. For a kind of response
(e.g., the response of AND type), however, modes of clustering
possibly depend on parameter values. For example, for a set of
parameter values given above, two kinds of clustering modes in the
case of AND have been found and displayed, but for a different set
of parameter values, other kinds of clustering modes are possible.
In addition, in the case that parameter values are chosen close to
the margin of oscillation of the uncoupled oscillator, the system
can display richer dynamical behaviors expecting to be further
investigated, but Kuramoto's phase reduction approach cannot be
used.

In addition, we point out that many theoretical studies have shown
that biological oscillators intertwined with positive and negative
feedback loops should have the following essential requirements
\cite{Tyson,Novak}. First, negative feedback is necessary to carry a
reaction network back to the `starting point' of its oscillation.
Second, the negative feedback signal must be sufficiently delayed in
time so that the chemical reactions do not settle on a stable steady
state. Third, the kinetic rate laws of the reaction mechanism must
be sufficiently `nonlinear' to destabilize the steady state. Fourth,
the reactions that produce and consume the interacting chemical
species must occur on appropriate timescales that permit the network
to generate oscillations. Facing to the complexity of gene
regulatory networks, these mathematical insights reveal the true
nature of gene relaxation oscillators. Our core relaxation
oscillator can show sustained and robust oscillation under the
guarantee of the above theoretical results. Especially, our coupled
positive and negative feedback biological oscillator models rely on
a separation of time scales between the two components to create
relaxation oscillations, i.e., the activator must have fast dynamics
than repressor. To that end, we can increase the plasmid copy number
concentrations as well as degradation rates of activator, where high
degradation rate has artificially been implemented by using peptide
sequences appended to the protein to make it a target for proteases
in the cell \cite{MBElowitz-Nature00,Gottesman}. Therefore, it would
be possible to experimentally demonstrate our circuit design. It
would be much more useful to take a hybrid approach in which
experiments and modeling can be performed in parallel to advance one
another. In a cyclic fashion, experiments can be used to inform the
designs of mathematical models, which can in turn be used to make
experimentally testable predictions.

Finally, ongoing structural, biochemical and cell-based studies have
begun to reveal several common principles by which protein
components are used to specifically transmit and process
information. Our studies demonstrate that these relatively simple
principles can be used to rewire signaling behaviors in a process
that mimics the evolution of new phenotypic responses. We expect
that our work would motivate the investigations in areas such as
development, where epigenetic inheritance leads to a persistent
phenotypic alteration in response to transient signals, or in
cell-cell communication systems that coordinate the rich complexity
of group behaviors.

\section*{ACKNOWLEDGMENT}

This work was supported by the Natural Science Key Foundation of
People's Republic of China (No. 60736028).

\appendix

\section{Okuda's approach}

In this appendix, we define cluster-balanced states and study their
stability. Each cluster contains the same number of oscillators.
Thus, we restrict our attention mainly on symmetric states.

Assume that the phase model of $N$ oscillators is governed by
\begin{eqnarray}
\frac{d\phi_i}{dt}=\Omega+\frac{1}{N}\sum^N_{j=1}\Gamma(\phi_i-\phi_j),
\end{eqnarray}
where $i=1,2,\cdots, N.$ Although $\Omega$ can be given any value in
a suitable moving coordinate, we assume $\Omega=0$ below without
explicitly refer to it. First, we define a symmetric $M$-cluster
state as the state in which $N/M$ oscillators belong to each of $M$
clusters. Since no randomness is including in the system, all the
oscillators in a certain cluster should be located at the same
phase. Let $\Phi_k$ denote the phase of cluster $k$
($k=0,1,\cdots,M-1$). From the phase equation, we obtain the
equation for $\Phi_k$ as
\begin{eqnarray}
\Phi_k=\frac{1}{M}\sum^{M-1}_{l=0}\Gamma(\Phi_k-\Phi_l).
\end{eqnarray}
We seek solutions to this equation in the form
\begin{eqnarray}
\Phi_k=\omega^{(M)}t+\frac{2\pi k}{M},
\end{eqnarray}
which implies that the phases of the $M$ clusters are equally
separated and rotate at a constant frequency $\omega^{(M)}$.
Substituting it into the above phase equation, we find that the
solution of the above form exists if
\begin{eqnarray}
\omega^{(M)}=\frac{1}{M}\sum^{M-1}_{k=0}\Gamma\left(\frac{2\pi
k}{M}\right).
\end{eqnarray}

Next, we analyze the stability of the balanced $M-$cluster state.
Let us put $\delta\phi_i=\phi_i-\Phi_k$ (where $i$ belongs to
cluster $k$) and express the linearized equation for $\delta\phi_i$
as $d\delta\Phi=J\Phi$, where the vector notation
$\delta\Phi=(\delta\phi_1,\delta\phi_ 2,\cdots,\delta\phi_N)$ and
$N\times N$ matrix $J$ have been used. Without loss of generality,
we assume that cluster $k$ consists of the oscillators with
$kN/M<i\leq(k+1)N/M$. Then , we have
\begin{eqnarray}
J=\left(\begin{array}{cccc}\alpha I-\beta_0 E & -\beta_1E & \cdots & -\beta_{M-1}E\\
-\beta_{M-1}E & \alpha I-\beta_0 E & \cdots & -\beta_{M-2}E\\
\cdots & \cdots & \cdots & \cdots\\
-\beta_1E & -\beta_2E & \cdots & \alpha I-\beta_0 E
\end{array}\right)
\end{eqnarray}
where $I$ is the $N/M\times N/M$ unit matrix and $E$ is a matrix of
the same dimension whose components are all $1$, $\alpha$ and
$\beta_k$ are expressed as
\begin{eqnarray}
\alpha=\frac{1}{M}\sum^{M-1}_{k=0}\Gamma'\left(\frac{2\pi
k}{M}\right),\,\,\,\, \beta_k=\frac{1}{N}\Gamma'\left(-\frac{2\pi
k}{M}\right),
\end{eqnarray}
and primes indicate the derivative with respect to the
argument. Since $J$ is a cyclic matrix, the explicit form of the
characteristic equation of $J$ can be obtained as
\begin{equation}
\begin{split}
|\lambda
I-J|&=\prod^{M-1}_{q=0}\left|(\lambda-\alpha)I+\left(\sum^{M-1}_{k=0}\beta_ke^{i2\pi
kq/M}E\right)\right|\\
&=(\lambda-\alpha)^{N-M}\prod^{M-1}_{q=0}\left(\lambda-\alpha+\frac{N}{M}
\sum^{M-1}_{k=0}\beta_ke^{i2\pi kq/M}\right)=0,
\end{split}
\end{equation}
where $i=\sqrt{-1}$. In this way, we obtain $N$ eigenvalues of $J$
in the form
\begin{eqnarray}
\lambda_p\equiv\alpha=\frac{1}{M}\sum^{M-1}_{k=0}\Gamma'\left(\frac{2\pi
k}{M}\right)\,\,\,\,(p=M,M+1,\,\,N-1),
\end{eqnarray}
\begin{equation}
\begin{split}
\lambda_q&\equiv\alpha-\frac{N}{M}\sum^{M-1}_{k=0}\beta_ke^{i2\pi
kq/M}\\
&=\frac{1}{M}\sum^{M-1}_{k=0}\Gamma'\left(\frac{2\pi
k}{M}\right)\left(1-e^{-i2\pi
kq/M}\right)\,\,\,\,(q=0,1,\cdots,M-1).
\end{split}
\end{equation}


\begin{thebibliography}{}
\bibitem{note}
The protein degradation rate can be experimentally altered by
augmenting or tagging the protein with addition amino.
\end{thebibliography}
\end{document}